\documentclass[pra,amsmath,preprint]{revtex4}

\usepackage{graphicx}

\begin{document}

\title{One-Electron Ionization of Multielectron Systems in Strong 
Nonresonant Laser Fields}

\author{Michael Spanner}
\affiliation{Steacie Institute for Molecular Sciences, 
National Research Council of Canada, Ottawa, ON, Canada K1A 0R6}
\author{Serguei Patchkovskii}
\affiliation{Steacie Institute for Molecular Sciences, 
National Research Council of Canada, Ottawa, ON, Canada K1A 0R6}

\begin{abstract}
We present a novel approach to calculating strong field ionization dynamics of
multielectron molecular targets.  Adopting a multielectron wavefunction ansatz
based on field-free {\it ab initio} neutral and ionic multielectron states, a
set of coupled time-dependent single-particle Schr\"odinger equations
describing the neutral amplitude and continuum electron are constructed.  These
equations, amenable to direct numerical solution or further analytical
treatment, allow one to study multielectron effects during strong field
ionization, recollision, and high harmonic generation.  We apply the method to
strong field ionization of CO$_2$, and suggest the importance of intermediate
core excitation to explain previous failure of analytical models to reproduce
experimental ionization yields for this molecule.  
\end{abstract}

\maketitle

\section{Introduction}

Present theoretical tools for calculating strong field ionization of atoms and
molecules fall into two categories, 1) semianalytical theories based the Strong
Field Approximation \cite{KFR} and/or ADK theory \cite{ADK}, often with
improvements over the traditional formulation to incorporate molecular targets
\cite{Becker,MOADK}, and 2) direct time-dependent numerical solution of the
Schr\"odinger equation.  The first category suffers from approximations
necessary to allow a semianalytical treatment, most notably the neglect of the
target-specific binding potential of the molecular core on the ionization, continuum, 
and recollision dynamics.  The second category has the shortcoming that full
numerical treatment becomes impossible as the number of degrees of freedom
increases.  Time-dependent numerical solutions of the Schr\"odinger equation
including a strong laser field is only feasible for one- or two-particle
systems.  Steps have been made along the numerical route to incorporate
multielectron effects into strong field dynamics through the use of
time-dependent Hartree-Fock theory \cite{TDHF}, multiconfigurational
time-dependent Hartree-Fock \cite{MCTDHF}, time-dependent configuration
interaction singles \cite{TDCIS}, and time-dependent density-functional theory
\cite{TDDFT}.  

In this work we address both the problems of including the binding potential
consistently throughout the strong field dynamics as well as the problem of
accounting for a major fraction of multielectron effects.  In particular,
motivated by recent experiments demonstrating effects of multiple final ionic
states in high harmonic generation (HHG) \cite{MultipleIonicStatesHHG}, we focus on a
multiple ionic channel effects in strong field ionization which is the first
step in HHG.  We consider only the electronic problem, with
the nuclei held fixed and work in the length gauge.  Our approach to strong
field ionization of multielectron targets combines {\it ab initio} quantum
chemistry multielectron wavefunctions with single particle time-dependent
numerical grid solutions.   We use as a basis the field-free $n$-electron
neutral and the lowest few ($n$-1)-electron singly ionized states.  Any
coupling to the multiply-charged ionic states is neglected.  The wavefunction
of the $n^{th}$ continuum electron associated with each ionic state is
represented by a 3D Cartesian numerical grid.  Equations of motion describing
the evolution and coupling of the basis state amplitudes and the $n^{th}$
electron wavefunction are derived from the multielectron Schr\"odinger equation
and contain no adjustable parameters.  Our method is closely related to the
R-matrix theory of electron-molecular scattering \cite{Rmatrix}.  We use the
identical wave function ansatz.  R-matrix theory accounts for
antisymmetrization exactly and is applicable to  time-independent problems
while our formalism includes antisymmetrization approximately but can be
applied to time-dependent problems.

As a first example, we apply the method to the strong field ionization of
CO$_2$.  A recent experiment \cite{Domagoj} found that predictions made using
MO-ADK for strong field ionization of CO$_2$ failed to account for the
experimental angle-resolved ionization yields.  Strong field ioniozation of
this molecule has also been theoretically analysed in recent papers using TDDFT
in Ref.\cite{ShihI} and single-channel frozen-core approach in
Ref.\cite{Madsen}.  Following our analysis presented below, we suggest that an
intermediate excitation channel not considered in Ref.\cite{Domagoj} is
responsible.  In this channel, first an excitation of the outer-lying electron
occurs concomitant with an ionic core excitation.  The excited ionic core then
couples back to the ground state of the inner core via laser coupling followed
by release of the outer-lying electron. 

\section{Length Gauge Theory for One-Electron Continuum}

\subsection{Hamiltonians and States}

The (non-relativistic) Hamiltonians of the laser-free ion and neutral are
\begin{equation}
	H^I(\{{\vec r}\}_{n-1}) =
	\sum_{i=1}^{n-1}\left[-\frac{1}{2}{\vec \nabla}^2_i + V_{nuc}({\vec r}_i)
             +\sum_{j=i+1}^{n-1}\frac{1}{|{\vec r}_i-{\vec r}_j|} \right]
\end{equation}
\begin{equation}
	H^N(\{{\vec r}\}_n) = H^I(\{{\vec r}\}_{n-1}) 
	-\frac{1}{2}{\vec \nabla}^2_n + V_{nuc}({\vec r}_n)
	                           +\sum_{i=1}^{n-1}\frac{1}{|{\vec r}_i-{\vec r}_n|}
\end{equation}
where $\{{\vec r}\}_{n-1}$ are the ($n$-1) spatial electronic coordinates of the 
ion, $\{{\vec r}\}_n$ are the $n$ spatial electronic coordinates of the
neutral, and $V_{nuc}({\vec r})$ is the electrostatic potential of the nuclei
\begin{equation}
	V_{nuc}({\vec r}) = \sum_a \frac{-Z_a}{|{\vec r}-{\vec R}_a|}
\end{equation}
where $Z_a$ and ${\vec R}_a$ are the charges and positions of the nuclei.
Note that Hartree atomic units ($\hbar = m_e = e = 1$) are used throughout.
In the length gauge and dipole approximation, the Hamiltonian of the full $n$ electron system 
interacting with the laser field is
\begin{equation}
	H^F(\{{\vec r}\}_n,t) = H^N(\{{\vec r}\}_n) - \sum_{i=1}^n {\vec F}(t) \cdot {\vec r}_i.
\end{equation}
Let $|N_j\rangle$ and $|I_j\rangle$ be the
orthogonal $n$-electron eigenstates of the field-free neutral 
and the ($n$-1)-electron eigenstates of the field-free ion respectively
\begin{eqnarray}\label{EqEigen}
	 H^N |N_j\rangle &=& E^N_j |N_j\rangle
	\\ \nonumber
	  H^I |I_j\rangle &=& E^I_j |I_j\rangle
\end{eqnarray}
Note that $|N_j\rangle$ and $|I_j\rangle$ depend on both spatial as well as spin coordinates 
of the electrons.
In practice, {\it ab initio} multielectron methods provide only approximate eigenstates.  The
approximate nature of $|N_j\rangle$ and $|I_j\rangle$ could be taken 
into account by using the expectation value equations 
\begin{eqnarray}
	\langle N_j | H^N |N_j\rangle &=& E^N_j 
	\\ \nonumber
	 \langle I_j| H^I |I_j\rangle &=& E^I_j 
\end{eqnarray}
instead of the eigenvalue equations Eqs.~(\ref{EqEigen}).
In this case,  whenever a term like $H^N|N_j\rangle$ is encountered in
the derivation, it must be replaced by the expansion
\begin{equation}
    H^N|N_j\rangle = \sum_i |N_i\rangle\langle N_i|H^N|N_j\rangle
\end{equation}
and likewise for the terms $H^I|I_j\rangle$.
Thus additional terms coupling the basis states $|N_j\rangle$  and $|I_j\rangle$ will arise that are 
not found in the formulation when Eqs.~(\ref{EqEigen}) hold.
For the present work it is assumed that the states are the exact neutral and ionic eigenstates
and Eqs.~(\ref{EqEigen}) are used in the following derivation.  
In the following only the neutral ground state $|N_0 \rangle = |N\rangle$ will be used.

\subsection{Antisymmetrization}

We use a wavefunction ansatz that has the form (see below for specific ansatz used)
\begin{equation}\label{EqProxy}
	|\Psi(t)\rangle = \widehat {\cal A} |\Psi_p(t)\rangle,
\end{equation}
where $|\Psi_p(t)\rangle$ is a non-antisymmetrized 'proxy' wavefunction ansatz that 
treats the $n^{th}$ electron differently than the remaining ($n$-1) core electrons,
\begin{equation}
	\widehat {\cal A} = \frac{1}{\sqrt{n}}\left( 1 - \sum_{j=1}^{n-1} \widehat P_{jn} \right)
\end{equation}
is the antisymmetrization operator that antisymmetrizes the $n^{th}$ electron with
the remaining ($n$-1) electrons, and $\widehat P_{jn}$ is the permutation
operator that interchanges electrons $j$ and $n$.  
Note that the ($n$-1) core electrons are already correctly antisymmetrized 
due to the use of fully antisymmetric $|N\rangle$ and $|I_m\rangle$ states.
If exact propagation of $n$-electron states were possible and if
the proxy wavefunction $|\Psi_p(t)\rangle$ spanned the full multi-electron space,
the time evolution of Eq.~(\ref{EqProxy}) would be given by
\begin{eqnarray}\label{NoSymmetry}
	\widehat U(t,t_0)|\Psi(t_0)\rangle 
	&=& \widehat U(t,t_0) \widehat {\cal A} |\Psi_p(t_0)\rangle
\\ \nonumber
	&=& \widehat U(t,t_0) \frac{1}{\sqrt{n}}\left( 1 - \sum_{j=1}^{n-1} \widehat P_{jn} \right)
	|\Psi_p(t_0)\rangle
\\ \nonumber
	&=&  \frac{1}{\sqrt{n}}\left( 1 - \sum_{j=1}^{n-1} \widehat P_{jn} \right)
	\widehat U(t,t_0) |\Psi_p(t_0)\rangle
\\ \nonumber
	&=&  \widehat {\cal A} \widehat U(t,t_0) |\Psi_p(t_0)\rangle,
\end{eqnarray}
where $\widehat U(t,t_0)$ is the evolution operator defined by
\begin{equation}
	i \frac{\partial}{\partial t}\widehat U(t,t_0) =  H^F(t) \widehat U(t,t_0),
	\:\:\: \widehat U(t_0,t_0) = \widehat I.
\end{equation}
Equation (\ref{NoSymmetry}) demonstrates that, at least in the case of
exact propagation, one need not propagate a fully antisymmetrized wavefunction.
Rather, it is enough to propagate a partially symmetrized initial state and
apply antisymmetrization at the final time: $\widehat {\cal A} \widehat U(t,t_0) |\Psi_p(t_0)\rangle$.

With this property of time evolution in mind, we proceed to construct a propagation scheme for
a non-antisymmetrized proxy wavefunction ansatz
\begin{equation}
	|\Psi_p(t)\rangle = \widehat U(t,t_0) |\Psi_p(t_0)\rangle
\end{equation}
where the $n^{th}$ electron is treated differently than the ($n$-1) core electrons.
The correctly antisymmetrized wavefunction can then be retrieved using Eq.~(\ref{EqProxy}).
Since the propagator construct below is only approximate, due to the use of
a truncated basis of ionic states, the reconstructed antisymmetric wavefunction will no longer
be an exact representation of time evolution of the initial antisymmetric wavefunction.
We will return to this point following the definition of $|\Psi_p(t)\rangle$ below 
to see what our propagation scheme missed using this procedure.

\subsection{Projectors and Wavefunction Ansatz}

We wish to construct a propagation scheme based on coupled
single-particle Schr\"odinger equations.  With this goal in mind,
we now introduce a set of single-particle orbitals that arise naturally for
the present problem, and the multi-electron partitioning that will be used below.  

Given the neutral ground state $|N\rangle$ and ionic states $|I_m\rangle$, we
introduce the set of (single-particle) orbitals, called ionization source
orbitals, defined as the overlap between the neutral and ionic states
\begin{equation}
	|\phi^S_m\rangle = \langle I_m | N\rangle
\end{equation}
where the integration is over the ($n$-1) electrons of the ion.
These source orbitals are related to the Dyson orbitals $|\psi^D_m\rangle$ that arise in 
photoionization processes \cite{Dyson1,Dyson2} by a simple 
scaling factor, $|\psi^D_m\rangle = \sqrt{n}|\phi^S_m\rangle$.
In addition, it will be convenient to use the {\it normalized} source orbitals 
$|\tilde \phi^S_m\rangle$, defined as
\begin{equation}
	|\tilde \phi^S_m\rangle = \frac{|\phi^S_m\rangle}{\sqrt{\langle \phi^S_m|\phi^S_m\rangle}},
\end{equation}
as well as the amplitude $\eta_m$:
\begin{equation}
	\eta_m = \langle \tilde \phi^S_m|\phi^S_m\rangle.
\end{equation}
Using $|\tilde \phi^S_m\rangle$ and its associated ionic states $|I_m\rangle$ we define
the multi-electron source-ion states $|S_m\rangle$ as
\begin{equation}
	|S_m\rangle = |\tilde \phi^S_m\rangle |I_m\rangle.
\end{equation}

We now introduce the set of projectors used below to partition the multi-electron wavefunction:
\begin{subequations}
\begin{eqnarray}
	\widehat {\cal P}^S_m &=& | S_m\rangle\langle S_m|
	\\ 
	\widehat {\cal P}^{\tilde N} &=& |\tilde N\rangle\langle\tilde N|
	\\ \nonumber
	&=& 
	\left( \widehat I - \sum_{k'} \widehat {\cal P}^S_{k'}\right)
	|N\rangle|{\cal N}_{\tilde N}|^2\langle N|
	\left( \widehat I - \sum_k \widehat {\cal P}^S_k\right)
	\\ 
	\widehat {\cal P}^I_m &=& 
	\left( \widehat I - \widehat {\cal P}^{\tilde N} - \sum_{k'} \widehat {\cal P}^S_{k'} \right)
	|I_m\rangle 
	\langle I_m|
	\left( \widehat I - \widehat {\cal P}^{\tilde N} - \sum_k \widehat {\cal P}^S_k \right)
	\nonumber \\
	&=& 
	\left( \widehat I - \widehat {\cal P}^{\tilde N} - \widehat {\cal P}^S_m \right)
	|I_m\rangle\langle I_m|
	\left( \widehat I - \widehat {\cal P}^{\tilde N} - \widehat {\cal P}^S_m \right)
\end{eqnarray}
\end{subequations}
where 
\begin{equation}
	|\tilde N\rangle = {\cal N}_{\tilde N} \left(\widehat I - \sum_m \widehat {\cal P}^S_m\right) |N\rangle
	= {\cal N}_{\tilde N}\Big[ |N\rangle - \sum_m \eta_m |S_m\rangle \Big]
\end{equation}
is the (normalized) component of the neutral ground state orthogonal to 
the set of source-ion states $|S_m\rangle$ used, and
\begin{equation}
	{\cal N}_{\tilde N} = \left(1-\sum_m|\eta_m|^2\right)^{-1/2}
\end{equation}
is the normalization factor of the state $|\tilde N\rangle$.  
These projectors split the multi-electron space into three parts with distinct physical interpretation: 
the $\widehat {\cal P}^S_m$ project onto the overlap between the neutral and ionic states,
$\widehat {\cal P}^{\tilde N}$ projects onto the component of the neutral 
that is orthogonal to all of the ionic states, and the $\widehat {\cal P}^I_m$ project
onto the component of the ionic channels that is orthogonal to the neutral.

The projectors defined above obey the standard relations
for a mutually orthogonal set of projectors
\begin{subequations}
\begin{eqnarray}
	\widehat {\cal P}^{\tilde N} \widehat {\cal P}^{\tilde N} &=& \widehat {\cal P}^{\tilde N}
	\\
	\widehat {\cal P}^S_m \widehat {\cal P}^S_k &=& \delta_{mk}\widehat {\cal P}^S_m
	\\
	\widehat {\cal P}^I_m \widehat {\cal P}^I_k &=& \delta_{mk} \widehat {\cal P}^I_m 
	\\ 
	\widehat {\cal P}^S_m \widehat {\cal P}^{\tilde N} &=&
	\widehat {\cal P}^{\tilde N} \widehat {\cal P}^S_m = 0
	\\
	\widehat {\cal P}^S_m \widehat {\cal P}^I_k &=& 
	\widehat {\cal P}^I_k \widehat {\cal P}^S_m = 0
	\\
	\widehat {\cal P}^{\tilde N} \widehat {\cal P}^I_m &=& 
	 \widehat {\cal P}^I_m \widehat {\cal P}^{\tilde N} = 0
\end{eqnarray}
\end{subequations}
where $\delta_{mk}$ is the Kronecker delta.
Further, Using these relations it can be shown that
\begin{equation}\label{EqRDIm}
	\langle I_m| \widehat {\cal P}^I_m = \widehat {\cal R}^S_m \langle I_m|,
\end{equation}
where $\widehat {\cal R}^S_m = (1-|\tilde\phi^S_m\rangle\langle\tilde\phi^S_m|)$ projects out
(removes) the source orbital from the one-particle space connected to the $|I_m\rangle$ channel. 
Equation (\ref{EqRDIm}) will be used below.

The wavefunction ansatz for the proxy wavefunction constructed in 
the space spanned by these projectors is
\begin{equation}\label{EqOvercomplete}
	|\Psi_p(t)\rangle = b(t)|\tilde N\rangle   +
	\sum_m\Big[ a_m(t)| S_m\rangle + |X_m(t)\rangle\Big]
\end{equation}
where 
\begin{equation}
	|X_m(t)\rangle = |\chi_m(t)\rangle|I_m\rangle
\end{equation}
and $|\chi_m(t)\rangle$ is the single-particle function that represents the
excited $n^{th}$ electron associated with the ionic channel $|I_m\rangle$, that
is, $|\chi_m(t)\rangle$ contains the continuum electron wavefunction that we
wish to calculate.  By imposing the condition $\langle S_m|X_m(t)\rangle =
\langle \phi^S_m|\chi_m(t)\rangle = 0$, which must be enforced in the initial
condition and is maintained during the propagation through the use of the
projection operators below, the basis states in $|\Psi_p(t)\rangle$ represent
orthogonal spaces that can be accessed by operating with the projection
operators
\begin{subequations}
\begin{eqnarray}
	\widehat {\cal P}^{\tilde N} |\Psi_p(t)\rangle &=& b(t) |\tilde N\rangle
	\\ 
	\widehat {\cal P}^S_m |\Psi_p(t)\rangle &=& a_m(t) | S_m\rangle
	\\ 
	\widehat {\cal P}^I_m |\Psi_p(t)\rangle &=& |\chi_m(t)\rangle|I_m\rangle.
\end{eqnarray}
\end{subequations}

Returning to the issue of antisymmetrization discussion in the previous
section, we can now point out the dominant interactions that are neglected
using the procedure
\begin{equation}\label{EqNoAntisym}
	\widehat U(t,t_0) \widehat {\cal A} |\Psi_p(t_0)\rangle
	\rightarrow  \widehat {\cal A} \widehat U(t,t_0)|\Psi_p(t_0)\rangle
\end{equation}
with the ansatz define in Eq.~(\ref{EqOvercomplete}).  First we note that by
using fully antisymmetric neutral $|N\rangle$ and ionic states $|I_m\rangle$,
correct antisymmetrization is present in the ($n$-1) core electrons.  Thus the
procedure in Eq.~(\ref{EqNoAntisym}) only concerns the $n^{th}$ (i.e.
continuum) electron.  When using a truncated basis of only a few low lying
$|I_m\rangle$ states, the representation given by Eq.~(\ref{EqOvercomplete})
only allows for a single electron (the $n^{th}$ electron) to be in highly
excited or continuum states.  Thus, no interactions that couple a continuum (or
highly excited) state of one electron with a continuum state of a {\it
different} electron are allowed in the present formulation.  Note that these
interactions are different than interactions of two electrons simultaneously in
the continuum, and would appear as two-particle operators that cause
transitions between two-electron states where, for example, a continuum state
of electron $j$ and a bound state of electron $k$ simultaneously couple to a
continuum states of electron $k$ and a bound state of electron $j$.

\begin{widetext}

\subsection{Full Propagation Equations}

Consider now the Schr\"odinger equation for $|\Psi_p(t)\rangle$
(where $\partial_t = \partial/\partial t$)
\begin{equation}
	i\partial_t |\Psi_p(t)\rangle = H^F(t) |\Psi_p(t)\rangle.
\end{equation}
The solution of this equation is equivalent to solving $\widehat U(t,t_0)|\Psi_p(t_0)\rangle$
discussed above.
Using the projection operators, the Schr\"odinger equation becomes
\begin{subequations}
\begin{eqnarray}
	  i \partial_t \widehat{\cal P}^{\tilde N}|\Psi_p(t)\rangle  
	&=& 
	 \widehat {\cal P}^{\tilde N}  H^F(t) \widehat {\cal P}^{\tilde N}|\Psi_p(t)\rangle
	+ \sum_k \widehat {\cal P}^{\tilde N}  H^F(t) \widehat {\cal P}^S_k|\Psi_p(t)\rangle 
	+ \sum_k \widehat {\cal P}^{\tilde N}  H^F(t) \widehat {\cal P}^I_k|\Psi_p(t)\rangle
	\\ 
	  i \partial_t \widehat{\cal P}^S_m|\Psi_p(t)\rangle  
	&=& 
	 \widehat {\cal P}^S_m  H^F(t) \widehat {\cal P}^{\tilde N}|\Psi_p(t)\rangle
	+ \sum_k \widehat {\cal P}^S_m  H^F(t) \widehat {\cal P}^S_k|\Psi_p(t)\rangle 
	+ \sum_k \widehat {\cal P}^S_m  H^F(t) \widehat {\cal P}^I_k|\Psi_p(t)\rangle
	\\ 
	  i \partial_t \widehat{\cal P}^I_m|\Psi_p(t)\rangle 
	&=& 
	\widehat {\cal P}^I_m  H^F(t) \widehat {\cal P}^{\tilde N}|\Psi_p(t)\rangle
	+ \sum_k \widehat {\cal P}^I_m  H^F(t) \widehat {\cal P}^S_k|\Psi_p(t)\rangle 
	+ \sum_k \widehat {\cal P}^I_m  H^F(t) \widehat {\cal P}^I_k|\Psi_p(t)\rangle.
\end{eqnarray}
\end{subequations}
By projecting out $|\tilde N\rangle$, $|S_m\rangle$, and $|I_m\rangle$, 
and recalling Eq.~(\ref{EqRDIm}), a coupled set of Schr\"odinger equations
for $a_m(t)$, $b(t)$ and $|\chi_m(t)\rangle$ is obtained
\begin{subequations}\label{EqFinalMultiIon}
\begin{eqnarray}
	  i \partial_t  b(t) 
	&=& 
	 \langle \tilde N|  H^F(t) |\tilde N\rangle b(t)
	+ \sum_k \langle \tilde N|  H^F(t) | S_k\rangle a_k(t)
	+ \sum_k \langle \tilde N|  H^F(t) |X_k(t)\rangle
	\\ 
	  i \partial_t  a_m(t) 
	&=& 
	 \langle S_m|  H^F(t) |\tilde N\rangle b(t)
	+ \sum_k \langle S_m|  H^F(t) | S_k\rangle a_k(t)
	+ \sum_k\langle S_m|  H^F(t) |X_k(t)\rangle
	\\ 
	  i \partial_t |\chi_m(t)\rangle 
	&=& 
	 \widehat {\cal R}^S_m \langle I_m|  H^F(t) |\tilde N\rangle b(t)
	+ \sum_k\widehat {\cal R}^S_m  \langle I_m|  H^F(t) | S_k\rangle a_k(t)
	+ \sum_k\widehat {\cal R}^S_m  \langle I_m|  H^F(t) |X_k(t)\rangle.
\end{eqnarray}
\end{subequations}
All the required matrix elements of $ H^F(t)$ are given in the Appendix.

The set of Eqs.~(\ref{EqFinalMultiIon}), together with the matrix elements
appearing in the Appendix, is the main result of this work.  In particular,
they allow for the use of coupled {\it single-particle} propagation methods to
solve for the $|\chi_m(t)\rangle$ wavefunctions rigorously coupled to the
multielectron states $|N\rangle$ and $|I_m\rangle$.  Furthermore, numerical
propagation of Eqs.~(\ref{EqFinalMultiIon}) does not involve non-local
potentials.

\end{widetext}

\section{Specific Cases and Numerical Results}

\subsection{Singlet Molecules with Uncoupled Ionic Channels}

Equations (\ref{EqFinalMultiIon}) are completely general and can be applied to
any target molecule regardless of symmetry or charge state.  In this section we
chose to consider the particular case of ionization from singlet molecules.
Further, for simplicity in the first implementation, we consider uncoupled
ionic channels.  That is, we consider ionization to multiple final ionic
states, but calculate ionization to each individually neglecting inter-channel
couplings.  

For ionization from a singlet closed-shell neutral to a particular final ion
state $|I_m\rangle$, the ion can be left in either spin-up or spin-down states.
Thus, with spin included, every final continuum-times-ion state has two
spin-related channels, $|I_m,\uparrow\rangle$ and $|I_m,\downarrow\rangle$,
each coupled to a continuum electron with opposite spin,
$|\chi_m(t),\downarrow\rangle$ and $|\chi_m(t),\uparrow\rangle$ respectively.
As long as any spin-orbit coupling is neglected, the two spin-related continuum
functions are identical in all respects except for the differing spin label.
In this case, the proxy wavefunction takes to form
\begin{eqnarray}\label{EqSinglet}
	|\Psi(t)\rangle &=& b(t)|\tilde N\rangle +  
	  \left[ a^\uparrow_m(t)| \tilde\phi^S_m,\uparrow \rangle  
	+ |\chi_m(t),\uparrow\rangle \right] |I_m,\downarrow\rangle
\\ \nonumber
	&& + \left[ a^\downarrow_m(t)| \tilde\phi^S_m,\downarrow \rangle  
	+ |\chi_m(t),\downarrow\rangle \right] |I_m,\uparrow\rangle,
\end{eqnarray}
and Eqs.~(\ref{EqFinalMultiIon}) reduces to
\begin{subequations}\label{EqSingletProp}
\begin{eqnarray}
	i\partial_t b(t) &=& {\cal H}^{\tilde N}(t) b(t) + 2 \langle {\cal T}_m|\phi^I_m(t)\rangle
	\\ 
	i\partial_t a_m(t) &=& \langle \tilde \phi^S_m| [|{\cal H}^I_m(t)\rangle + b(t) |{\cal T}_m\rangle]
	\\
	i\partial_t |\chi_m(t)\rangle &=& \widehat {\cal R}^S_m [|{\cal H}^I_m(t)\rangle + b(t) |{\cal T}_m\rangle]
\end{eqnarray}
\end{subequations}
where $a_m(t) = a^\uparrow_m(t) = a^\downarrow_m(t)$, $|\chi_m(t)\rangle$
represents the (identical) spatial part of $|\chi_m(t),\downarrow\rangle$ and
$|\chi_m(t),\uparrow\rangle$, and 
\begin{equation}
	|\phi^I_m(t)\rangle = |\chi_m(t)\rangle + a_m(t)|\tilde \phi^S_m\rangle
\end{equation}
where $|\tilde \phi^S_m\rangle$ is the (identical) spatial part of the two
spin-related source orbitals $| \tilde\phi^S_m,\uparrow \rangle$ and $|
\tilde\phi^S_m,\downarrow \rangle$.  (In the following, we drop the explicit
spin dependence of the states when the quantities involved to do not dependent
on the spin label.) Also appearing in Eqs.~(\ref{EqSingletProp}) are
\begin{equation}
	|{\cal H}^I_m(t)\rangle = [ H_m - \vec F(t)\cdot(\vec r_n-\vec d^I_{mm})] |\phi^I_m(t)\rangle,
\end{equation}
where 
\begin{equation}
	H_m(\vec r_n) =  E^I_m-\frac{1}{2}{\vec \nabla}^2_n + V_{nuc}({\vec r}_n)
	+ V^H_{mm}({\vec r}_n)  ,
\end{equation}
is the single-electron field-free Hamiltonian for the $n^{th}$ electron 
moving in the field of the $m^{th}$ ionic state,
\begin{equation}
	{\vec d}^I_{mm} = -\langle I_m| \sum_{k=1}^{n-1} \vec r_k  |I_m\rangle
\end{equation}
is the electronic dipole moment of the ion, 
\begin{equation}
	V^H_{mm}(\vec r_n) = \langle I_m| \sum_{k=1}^{n-1}\frac{1}{|{\vec r}_k-{\vec r}_n|} |I_m\rangle
\end{equation}
is the electrostatic potential of the ion core electrons.
The (single particle) orbital $|{\cal T}^{\tilde N}_m\rangle$ defined as
\begin{equation}
	|{\cal T}^{\tilde N}_m\rangle = {\cal N}_{\tilde N} \Big[ 
	\eta_m[E^N_0-\vec F(t)\cdot \vec d^I_{mm} -  H_m]
	|\tilde \phi^S_m\rangle - \vec F(t)\cdot |\vec\phi^C_m\rangle \Big]
\end{equation}
is the 'transfer orbital' that couples $|\tilde\phi^S_m\rangle|I_m\rangle$ and 
$|\chi_m(t)\rangle|I_m\rangle$ to the $|\tilde N\rangle$ component of the neutral,  
where $|\vec\phi^C_m\rangle$ is given by
\begin{equation}
	|\vec\phi^C_m\rangle = \langle I_m| \sum_{k=1}^{n-1} \vec r_k  |N\rangle.
\end{equation}
This single-particle function $|\vec\phi^C_m\rangle$ represents an ionization
(or excitation) process where the laser field acts on a bound electron, but
ionizes (or excites) a different electron.  We refer to this orbital as a
'cradle orbital' in analogy with Newton's cradle, a multi-ball pendulum where
one ball receives a force causing a different ball to swing.  The remaining
term in Eqs.~(\ref{EqSingletProp}) given by
\begin{eqnarray}
	{\cal H}^{\tilde N}(t) &=& |{\cal N}_{\tilde N}|^2 \Big\{ E^N_0 + 2 |\eta_m|^2
	[\langle\tilde \phi^S_m| H_m|\tilde \phi^S_m\rangle -2 E^N_0]
	\nonumber \\
	&+& \vec F(t)\cdot \Big[ 
	\vec d^N + 2 |\eta_m|^2\vec d^I_{mm} 
	+ 2 |\eta_m|^2 \langle\tilde \phi^S_m|\vec r_n |\tilde \phi^S_m\rangle 
	\nonumber \\
	&+& 2 \eta_m\langle \vec \phi^C_m|\tilde \phi^S_m\rangle 
	 +  2 \eta^*_m\langle \tilde \phi^S_m|\vec \phi^C_m\rangle 
	\Big]
	\Big\}
\end{eqnarray}
is the energy of the $|\tilde N\rangle$ state in the presence of the laser field,
and 
\begin{equation}
    {\vec d}^N = -\langle N| \sum_{k=1}^n \vec r_k  |N\rangle
\end{equation}
is the electronic dipole moment of the neutral.
The initial condition corresponding to all population in the neutral state are
\begin{subequations}
\begin{eqnarray}
	b(t=0) &=& \sqrt{1-2 |\eta_m|^2}
	\\
	a_m(t=0) &=& \eta_m
	\\
	|\chi_m(t=0)\rangle &=& 0
\end{eqnarray}
\end{subequations}

The propagation equations (\ref{EqSingletProp}) coupling the continuum electron
$|\chi_m(t)\rangle$ to the ground state amplitudes $a_m(t)$ and $b(t)$ are
perhaps not so transparent at first glance.  They can be simplified in the case
of negligible depletion and distortion of the ground state,
\begin{subequations}
\begin{eqnarray}
	b(t) &\approx& b(t=0) e^{-iE^N_0(t) t}  \\
	a_m(t) &\approx& a_m(t=0) e^{-iE^N_0(t) t}, 
\end{eqnarray}
\end{subequations}
where $E^N_0(t) = E^N_0 - \vec F(t)\cdot \vec d^N - \alpha |\vec F(t)|^2$ 
takes into account a small Stark shift of the neutral.
In this case, Eq. (\ref{EqSingletProp}c) simplifies to 
\begin{eqnarray}\label{EqSingletApprox}
	i\partial_t |\chi_m(t)\rangle &=& \widehat {\cal R}^S_m \Big\{
	[ H_m - \vec F(t)\cdot(\vec r_n-\vec d^I_{mm})] |\chi_m(t)\rangle \Big\}
	\\ \nonumber
	&+& \widehat {\cal R}^S_m \Big\{ -\vec F(t)\cdot [\vec r |\phi^S_m\rangle + |\vec \phi^C_m\rangle] \Big\}
	 e^{-iE^N_0(t)t}.
\end{eqnarray}
This last equation is now very close to a standard laser-dressed
single-particle Schr\"odinger equation for $|\chi_m(t)\rangle$.  The only
difference is that orthogonality with the neutral is maintained through the
appearance of $\widehat {\cal R}^S_m$, and the term $\widehat {\cal R}^S_m \{
-\vec F(t)\cdot [\vec r |\phi^S_m\rangle + |\vec \phi^C_m\rangle] \}$ acts as
the source that populates $|\chi_m(t)\rangle$.  For regimes where negligible
depletion is expected and where Stark shifts and distortions of the neutral are
small, Eqn.~(\ref{EqSingletApprox}) could be used instead of
Eqs.~(\ref{EqSingletProp}).  In the following calculations, we use
Eqs.~(\ref{EqSingletProp}) throughout.

\subsection{Ionization of CO$_2$}

\begin{table}[b]
	\caption{Multielectron states and energies used in the CO$_2$ ionization calculations. 
	Zero of energy was set equal to the (degenerate) ionic ground state.}
	\label{TabStatesEnergies}
	\begin{ruledtabular}
	\begin{tabular}{cccc}
	State & Label & Energy (eV) & Hole\\
	\hline
	$|N\rangle$ & $\widetilde {\rm X}^1\Sigma_g$ & -13.76 \\
	$|I_1\rangle$ & $\widetilde {\rm X}^2\Pi_{g,x}$ & 0 & HOMO \\
	$|I_2\rangle$ & $\widetilde {\rm X}^2\Pi_{g,y}$ & 0 & HOMO \\
	$|I_3\rangle$ & $\widetilde {\rm A}^2\Pi_{u,x}$ & 3.53 & HOMO-1 \\
	$|I_4\rangle$ & $\widetilde {\rm A}^2\Pi_{u,y}$ & 3.53 & HOMO-1 \\
	$|I_5\rangle$ & $\widetilde {\rm B}^2\Sigma_u$ & 4.28 & HOMO-2
	\end{tabular}
	\end{ruledtabular}
\end{table}

We now apply this formalism to the strong field ionization of CO$_2$.
Recently, angle-resolved ionization yields have been measured \cite{Domagoj}
for this molecule, where the angle is between the molecular axis and the
polarization direction of a linearly polarized laser field.  In
Ref.~\cite{Domagoj} it was found that the experimental angular ionization
pattern for CO$_2$ differs strongly from the results of molecular ADK theory
(MO-ADK), a single-active electron quasi-static tunneling theory of molecular
ionization \cite{MOADK}.  The central difference is that MO-ADK predicts
ionization peaks at an angle of $\sim 30^o$ while the measure show strong peaks
at $\sim 45^o$.  Here we consider angle-resolved ionization yields of CO$_2$
exposed to a single cycle of an 800 nm laser ($\omega$ = 0.057 a.u.).

\begin{figure}[t]
	\centering
 	\includegraphics[width=0.5\columnwidth]{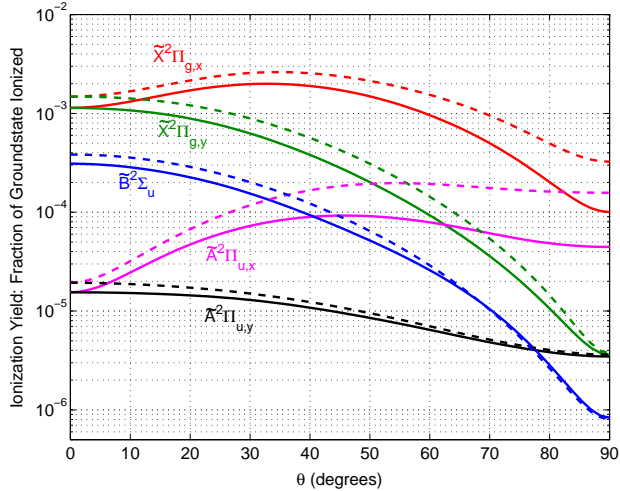}
	\caption{(Color online) Angle-resolved ionization yields for intensity 
	$1.5\times 10^{14}$ W/cm$^2$.  $\theta$ is 
	angle between the molecular axis and the polarization axis of
	the laser field.  The solid curves are results using a grid spacing $\Delta$ = 0.1 a.u.
	while the dashed curves used $\Delta$ = 0.2 a.u. }
	\label{FigYields}
\end{figure}

\begin{figure}[t]
	\centering
 	\includegraphics[angle=-90,width=0.8\columnwidth]{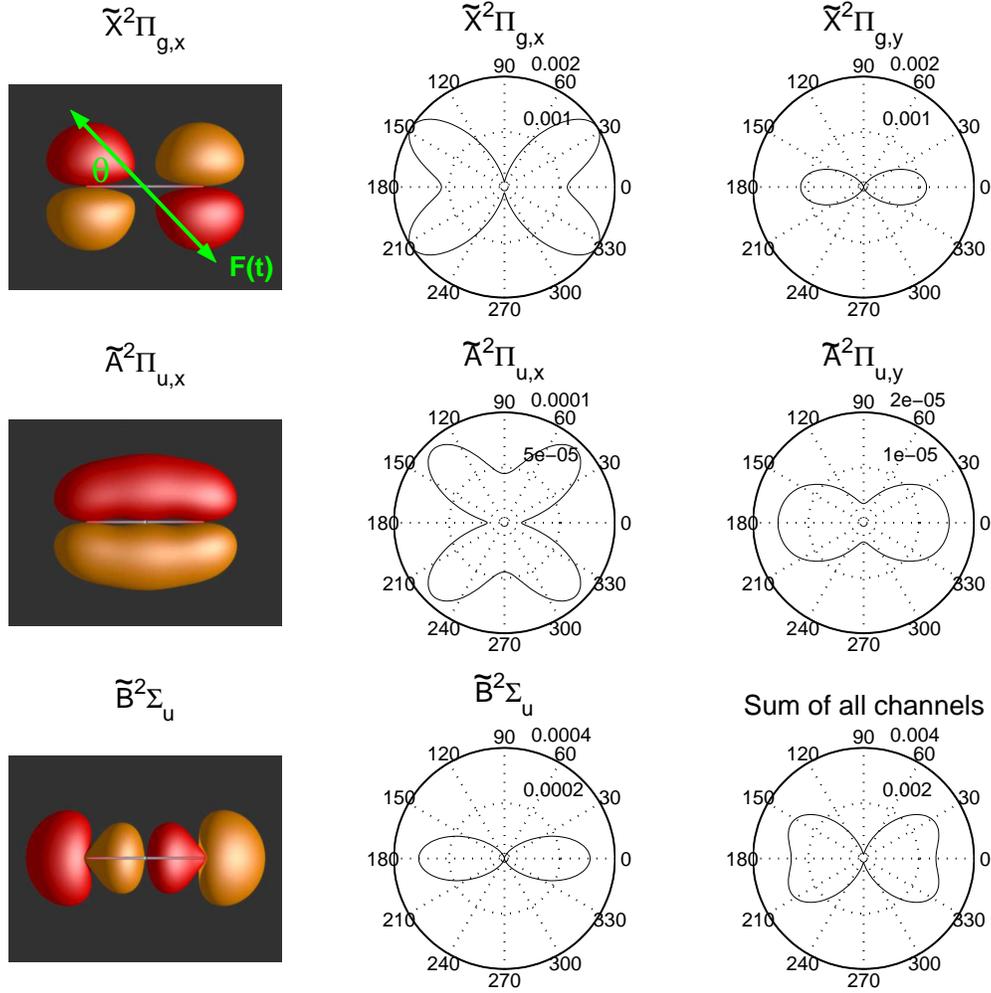}
	\caption{(Color online) Left column: Dyson orbitals of the ionic channels considered.  Angle between the 
	laser field and the molecular axis is depicted in the top left panel.  Only the Dyson orbitals
	with lobes in the plane of the laser field are shown.  
	Center and right columns: Angular ionization yields for intensity
    $1.5\times 10^{14}$ W/cm$^2$ for each ionic channel considered.
	Bottom-right panel shows the total ionization yield which is the sum of all channels.}
	\label{FigPolar}
\end{figure}

The neutral $|N\rangle$ and lowest five ionic $|I_m\rangle$ multielectron
orbitals are calculated using the GAMESS electronic structure code
\cite{GAMESS}.  All calculations use the cc-pVTZ basis set \cite{ccpVTZ} and
were done at a CAS level using 16 (neutral) or 15 (cation) active electrons in
10 orbitals.  Experimental geometry of the CO$_2$ ground state is used (linear,
$R_{\mathrm{C-O}}$=1.1621 $\AA$).  The states and energies used are shown in
Table~\ref{TabStatesEnergies} along with the approximate location of the hole
(relative to the neutral) left by the removed electron for each ionic state.
Equations (\ref{EqSinglet}) are solved using a leapfrog algorithm.  The
wavefunction $|\chi_m(t)\rangle$ is represented on a 3-dimensional Cartesian
grid.  The grid extends to $\pm$ 13 a.u. in the x and z directions, and to
$\pm$ 8 a.u. in the y direction.  All calculations are done with a grid spacing of
$\Delta$ = 0.1 a.u. in all directions unless otherwise specified.  Absorbing
boundary conditions are used in the xz plane with a width of 5 a.u.  from the
boundary edges \cite{Absorb}.  The ionization yield was calculated by
monitoring the density absorbed at the boundaries.  The CO$_2$ molecule has the
C atom at the origin and has the bond axis aligned along the z-axis.  The laser
field $F(t) = {\cal E}_0\sin(\omega t)$ is rotated in the xz plane. The angle
$\theta$ is the angle between the laser polarization and the molecular axis.
The laser field is turned off after a single cycle, $F$($t>2\pi/\omega$) = 0,
and the simulations are run until $t$ = 150 fs (an additional 40 fs after the
single cycle is over) to allow the liberated electron density to be absorbed at
the boundary.  A time step of $\Delta t$ = 0.00133 a.u. is used for the time
propagation.

Figure \ref{FigYields} plots angle-resolved ionization yields for the five
final ion states considered for a intensity of $1.5\times 10^{14}$ W/cm$^2$.
The solid lines correspond to calculations with the step sizes specified above,
while the dashed lines show results using $\Delta$ = 0.2 and $\Delta t$ =
0.00266 a.u.  While the total yields continue to decrease a bit as the grid
size becomes finer, the general character and relative behavior of the
ionization channels is preserved.  For all angles, the ionization yield is
dominated by the $\widetilde {\rm X}^2\Pi_g$ channels.  Polar plots showing the
angular shape of each ionization channel are presented in Fig.~\ref{FigPolar}.
Also shown in this plot is the total ionization yield that included the yield
from all channels (bottom-right panel), which is effectively the same as the
yield including only the two $\widetilde {\rm X}^2\Pi_g$ channels (not shown).
The total ionization yield has a 'bow tie'-like pattern, with peak values
appearing near 30$^o$.  This is in closer agreement with the MO-ADK results
than the experimental distributions, both presented in Ref.~\cite{Domagoj}.
Note that the MO-ADK results of  Ref.~\cite{Domagoj} include only the
`in-plane' HOMO channel which would correspond to the $\widetilde {\rm
X}^2\Pi_{g,x}$ channel alone.  Thus, our uncoupled channel calculations still
fail to reproduce the experimental peak positions seen in Ref.~\cite{Domagoj}.

\subsection{Role of Nodal Planes and the Binding Potential}

It has been shown that the presence of nodal planes in the ionizing orbitals
leads to suppression of the ionization rate \cite{Becker}.  Most prominently,
large suppression is expected to occur when the laser field is aligned along a
nodal plane.  This expected trend can be seen in our results
(Fig.~\ref{FigPolar}) by comparing the angular ionization yields with the
corresponding Dyson orbitals.  However, two features stand out that deserve
attention.  First, although suppression is seen along both nodal planes in the
$\widehat {\rm X}^2\Pi_{g,x}$ distribution, there is much more suppression
along the 90$^o$ node than along the $0^o$ node.  Second, there is a dip in the
$\widehat {\rm A}^2\Pi_{u,x}$ ionization yield at 90$^o$ that corresponds to no
obvious feature in the $\widehat {\rm A}^2\Pi_{u,x}$ Dyson orbital. 

\begin{figure}[t]
	\centering
 	\includegraphics[width=0.5\columnwidth]{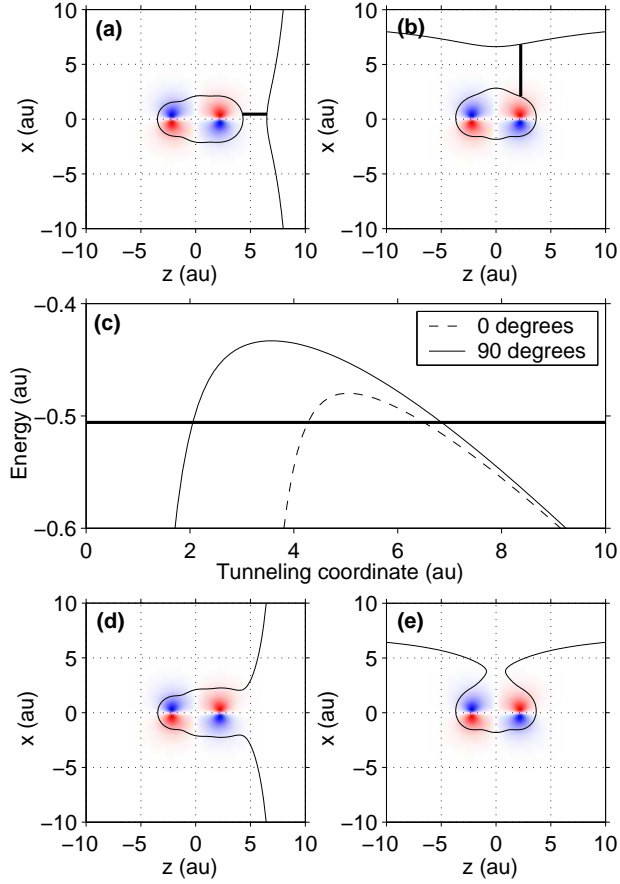}
	\caption{(Color online) Top row: Dyson orbital of the $\widehat {\rm X}^2\Pi_{g,x}$ channel 
	and ground state energy contours of
	the potential energy landscape at the peak of the laser field for $1\times10^{14}$ W/cm$^2$.
	In (a) the laser points along 0$^o$ while in (b) the laser field points along 90$^o$.
	Also shown are the tunneling paths through which the orbitals lobes must tunnel.
	Panel (c) plots the potential energy along the tunneling paths.  The solid line
	denotes the ground state energy of -13.76 eV = -0.5058 a.u.  Panels (d) and
	(e) plot the $\widehat {\rm X}^2\Pi_{g,x}$ Dyson orbital along with the ground state energy
	contours of the potential energy for $1.5\times10^{14}$ W/cm$^2$.  In panel (d)
	the laser field points along 0$^o$ while in panel (e) the laser field points along 90$^o$.
	}
	\label{Fig1B1contours}
\end{figure}

Consider the $\widehat {\rm X}^2\Pi_{g,x}$ distribution.  We first consider the
case when the peak laser field is $1\times10^{14}$ W/cm$^2$ and return to the
case of $1.5\times10^{14}$ W/cm$^2$ below.  (Note that although the angular
ionization yields shown in Fig.~\ref{FigPolar} where calculated for
$1.5\times10^{14}$ W/cm$^2$, the angular shapes for each channel are very
similar when using an intensity of $1\times10^{14}$ W/cm$^2$.) Panels (a) and
(b) in Fig.~\ref{Fig1B1contours} plot the $\widehat {\rm X}^2\Pi_{g,x}$ Dyson
orbital along with select contours of the instantaneous potential at the peak
of the laser pulse for an intensity of $1\times10^{14}$ W/cm$^2$.  The contours
are taken at the ground state energy of the neutral and show the entrance and
exit of the tunneling barrier through which the $\widehat {\rm X}^2\Pi_{g,x}$
Dyson orbital must escape.  Panel (a) shows the contours when the laser is
aligned along 0$^o$, while panel (b) is for the 90$^o$ case.  The short solid
lines connecting the entrance and exit depict the tunneling path positioned
along the peak of the orbitals lobes.  The tunneling barrier along these paths
are shown in panel (c).  Already one can see that the tunneling path through
which the orbital lobes must pass is much shorter in the 0$^o$ configuration
than in the 90$^o$ configuration suggesting the origin of the difference in
suppression in the 0$^o$ and 90$^o$ degrees directions seen in the $\widehat
{\rm X}^2\Pi_{g,x}$ distribution.  In order to get a quantitative semiclassical
estimate of the ratio of ionization at 0$^o$ and 90$^o$, we use the WKB
tunneling formula
\begin{equation}
	{\rm Rate} \sim \exp\left[-2\int^{x_1}_{x_0} \sqrt{2(V(x')-E)}dx'\right]
\end{equation}
where the integral is taken across the tunneling barrier.  Using this measure,
we find that the rate of tunneling along 0$^o$ should be larger than the rate
along 90$^o$ by a factor of 7.3, which in good agreement with the actual ratio
of 8.8 extracted from the simulations.  Panels (d) and (e) plot the same
contours as in panels (a) and (b), but now for the intensity of
$1.5\times10^{14}$ W/cm$^2$.  In this case, the ionization is above barrier,
and the ground state energy contours show the `doorway' opened by the presence
of the laser field.  Although a quantitative estimate is difficult in the
above-barrier regime, one can see that for 0$^o$ the doorway encompasses almost
the whole width of the Dyson orbital along this direction, while for 90$^o$ the
doorway is allowing only a small portion of the orbital localized around the
nodal plane to pass.  Thus, the analysis of the potential landscape in the
$1.5\times10^{14}$ W/cm$^2$ case allows for a qualitative understanding of the
large difference in suppression along the nodal planes seen in the $\widehat
{\rm X}^2\Pi_{g,x}$ angular ionization yields.

\begin{figure}[t]
	\centering
 	\includegraphics[width=0.5\columnwidth]{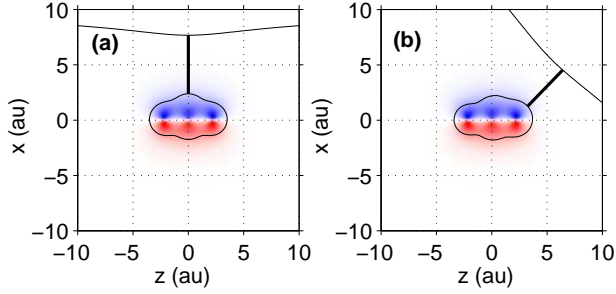}
	\caption{(Color online) Dyson orbital of the $\widehat {\rm A}^2\Pi_{u,x}$ channel 
	and ground state energy contours of
	the corresponding potential energy landscape at the peak of the 
	laser field for $1.5\times10^{14}$ W/cm$^2$.
	In (a) the laser points along 90$^o$ while in (b) the laser field points along 45$^o$.
	The thick line segments lines show the shortest the tunneling paths.
	}
	\label{Fig2B1contours}
\end{figure}

We turn now to the $\widehat {\rm A}^2\Pi_{u,x}$ channel, where a similar
analysis accounts for the dip at 90$^o$.  Figure \ref{Fig2B1contours} shows the
$\widehat {\rm A}^2\Pi_{u,x}$ Dyson orbital along with the ground state energy
contours.  In panel (a) the laser field points along 90$^o$ while panel (b)
corresponds to 45$^o$.  Both panels correspond to a peak intensity of
$1.5\times10^{14}$ W/cm$^2$, the case shown in Fig.~\ref{FigPolar}.
Integrating the tunneling rate along the paths shown in the plots, which are
the shortest paths connecting the inner and outer regions in both cases, we
calculated that the tunneling rate for the 90$^o$ case should be suppressed by
a factor of 0.6 as compared to the 45$^o$ case.  This is again in good
agreement with the actual suppression of 0.5 extracted from the results in
Fig.~\ref{FigPolar} for this channel.

\subsection{Toward Coupled-Channel Ionization of CO$_2$}

We can use the results of the present 
uncoupled channel calculations to infer potentially important 
ionization mechanisms that will appear in a coupled channel 
treatment.  

\begin{figure}[b]
	\centering
 	\includegraphics[width=0.5\columnwidth]{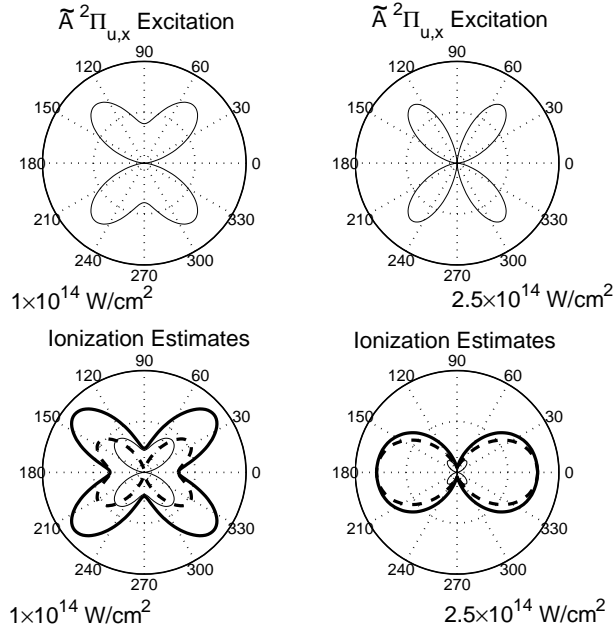}
	\caption{Top panels: Excitation yield surrounding the 
	$\widetilde {\rm A}^2\Pi_{u,x}$ ionic core for the two intensities
	of $1\times 10^{14}$ W/cm$^2$ and $2.5\times 10^{14}$ W/cm$^2$.
	Bottom panels: Ionization estimates for the two intensities 
	showing the direct ionization channel (thick-dashed) analogous to the
	"sum over all channels" panel shown in Fig.2,
	the estimated intermediate excitation channel (thin), 
	and the sum of these two channels (thick).}
	\label{FigExPolar}
\end{figure}

\begin{figure}[t]
	\centering
 	\includegraphics[width=0.5\columnwidth]{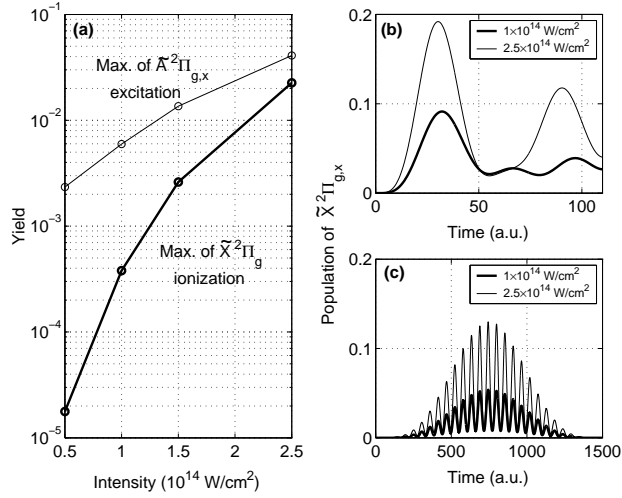}
	\caption{Panel (a): Maximum of $\widetilde {\rm A}^2\Pi_{u,x}$ 
	excitation and maximum of $\widetilde {\rm X}^2\Pi_g$ ionization
	as intensity is varied.  Panels (b) and (c) plot the population of the $\widetilde {\rm X}^2\Pi_g$
	ionic state using the 2-level laser coupled model of Eq.~(\ref{EqTwoLevel}), with
	all population initially in the $\widetilde {\rm A}^2\Pi_{u,x}$ state.  
	Panel (b) is for a single cycle pulse,
	and panel (c) is for a longer pulse with smooth Gaussian envelope.  }
	\label{FigExYields}
\end{figure}

In our formulation, the wavefunction $|\chi_m(t)\rangle$ carries not only the
continuum states, but also a complete set of bound states bound to the
$|I_m\rangle$ ionic core.  Thus, using the same simulations discussed above, we
can calculate excited, but un-ionized, population of $n^{th}$ electron
surrounding each ionic core.  In particular, the top two panels of
Fig.~\ref{FigExPolar} show the angular excitation yields surrounding the
$\widetilde {\rm A}^2\Pi_{u,x}$ ionic core for two intensities of 1$\times
10^{14}$ and 2.5$\times 10^{14}$ W/cm$^2$.  These yields show strong peaks near
(or beyond) 45$^o$.  In addition, as shown in Fig.~\ref{FigExYields}(a), the
peak excitation yield surrounding the $\widetilde {\rm A}^2\Pi_{u,x}$ ionic
core is much larger than the peak ionization yield coming from the $\widetilde
{\rm X}^2\Pi_g$ channels.  

In an uncoupled channel formulation, as is the case with the present
calculations, this excited population surrounding the $\widetilde {\rm
A}^2\Pi_{u,x}$ ionic core is trapped.  (We have checked that similar excited
population exists at the end of a 5 fs Gaussian laser pulse in addition to the
single cycle pulses used herein).  However, in a coupled channel formulation,
some of this excited population surrounding the $\widetilde {\rm A}^2\Pi_{u,x}$
core will be moved to the $\widetilde {\rm X}^2\Pi_{g,x}$ ionic core through
laser-induced dipole coupling of the $\widetilde {\rm A}^2\Pi_{u,x}$ and
$\widetilde {\rm X}^2\Pi_{g,x}$ core, i.e.  through the polarization of the
ionic cores.  The amount of ionic core coupling can be estimated by solving a
2-state problem for the laser coupling of the $\widetilde {\rm A}^2\Pi_{u,x}$
and $\widetilde {\rm X}^2\Pi_{g,x}$ cores
\begin{equation}\label{EqTwoLevel}
	i\frac{\partial}{\partial t}
	\left[ \begin{array}{c} C_X(t) \\ C_A(t) \end{array} \right] 
	= \begin{bmatrix} E_A & -F(t)\mu_{AB} \\ -F(t)\mu_{AB} & E_B \end{bmatrix} 
	\left[ \begin{array}{c} C_X(t) \\ C_A(t) \end{array} \right]
\end{equation}
where $\mu_{AB}$ = -0.46722 a.u. is the transition dipole between the ionic
states $\widetilde {\rm X}^2\Pi_{g,x}$ and $\widetilde {\rm A}^2\Pi_{u,x}$,
calculated using GAMESS as outlined above, and $C_X(t)$ and $C_A(t)$ are
amplitudes of the $\widetilde {\rm X}^2\Pi_{g,x}$ and $\widetilde {\rm
A}^2\Pi_{u,x}$ states.  Figures \ref{FigExYields} (b) and (c) plot $|C_X(t)|$
as a functions of time for two different pulse, panel (b) uses a single cycle
and panel (c) uses a multi-cycle pulse with Gaussian envelope, with the initial
condition $C_X(t)=0$ and $C_A(t)=1$.  The calculations were done for two
different intensities, $1\times 10^{14}$ W/cm$^2$ (thick lines) and $2.5\times
10^{14}$ W/cm$^2$ (thin lines).  These calculations allows us to estimate that
about 5 to 10\% of the excited population surrounding the $\widetilde {\rm
A}^2\Pi_{u,x}$ core will couple back to the $\widetilde {\rm X}^2\Pi_{g,x}$
state on subsequent cycles.  Some (and perhaps all) of this excited population
will escape the core region once coupled back to the $\widetilde {\rm
X}^2\Pi_{g,x}$ ionic core.  We thus anticipate two important ionization
channels in a coupled-channel formulation of CO$_2$, the direct channel and an
intermediate excitation channel
\begin{eqnarray}
	{\rm Direct:} & \:\:\: {\rm CO}_2(\widetilde{\rm X}^1\Sigma_g) 
	\rightarrow {\rm CO}^+_2(\widetilde{\rm X}^2\Sigma_g) + e^-
\\
	{\rm Inter.\: Ex.:} & \:\:\: {\rm CO}_2(\widetilde{\rm X}^1\Sigma_g) 
	\rightarrow {\rm CO}^+_2(\widetilde{\rm A}^2\Sigma_{u,x}) (e^-)^*
	 \rightarrow {\rm CO}^+_2(\widetilde{\rm X}^2\Sigma_g) + e^-
\end{eqnarray}
where $(e^-)^*$ denotes an excited electron.  Assuming that all of the excited
population will escape the core upon coupling from the $\widetilde {\rm
A}^2\Pi_{u,x}$ back to the $\widetilde {\rm X}^2\Pi_{g,x}$ state, the
intermediate excitation channel will carry predominantly the angular imprint of
the ${\rm CO}_2(\widetilde{\rm X}^1\Sigma_g) \longrightarrow {\rm
CO}^+_2(\widetilde{\rm A}^2\Sigma_{u,x}) (e^-)^*$ excitation step.  

The direct channel yield and (estimated) intermediate channel yield, as well as
their sum, is plotted in the bottom two panels of Fig.~\ref{FigExPolar} for the
two intensities shown.  Here the intermediate excitation channel yield was
estimated by multiplying the yields for excitation on the $\widetilde {\rm
A}^2\Pi_{u,x}$ ionic core by the amount of coupling seen in
Figs.~\ref{FigExYields}(b) and (c), 0.05 in the case of $1\times 10^{14}$
W/cm$^2$ and 0.1 for $2.5\times 10^{14}$ W/cm$^2$.  At the higher intensity,
the direct channel dominates, while for the lower intensity the intermediate
excitation channel is becoming important.  Further, the peak of the total
ionization estimate for $1\times 10^{14}$ W/cm$^2$ is now approaching 45$^o$,
as seen in the experiment \cite{Domagoj}.  Our treatment of the proposed
intermediate excitation channel is admittedly crude, and fails to reproduce the
sharpness of the experimental peaks seen in Ref.\cite{Domagoj}.  An accurate
description requires a full coupled-channel treatment of
Eqs.~(\ref{EqFinalMultiIon}) that includes at least the $\widetilde {\rm
X}^2\Pi_{g,x}$ and $\widetilde {\rm A}^2\Pi_{u,x}$ states.  However, from the
scaling of the excitation and ionization yields seen in
Fig.~\ref{FigExYields}(a) it is clear that the intermediate excitation channel
will become important for a correct description of strong field ionization of
CO$_2$ at intensities up to (and perhaps beyond) $10^{14}$ W/cm$^2$.

\section{Summary}

In this work we developed a method for strong field one-electron ionization of
multielectron targets.  Our method uses field-free multielectron orbitals to
describe the neutral and lowest few ionic states.  These multielectron basis
states are supplemented with a one-particle numerical grid used to represent
the continuum electron.  Equations of motion coupling the basis states to the
continuum grid are derived from the multielectron Schr\"odinger equation.  The
result is a coupled set of single-particle Schr\"odinger equations describing
ionization into each final ion state included in the ionic basis.  Our
equations are general and applicable to strong field ionization of any small
molecule.

As an example, we studied ionization of CO$_2$ in the uncoupled channel
approximation including the lowest five ionic states of CO$^+_2$.  Strong field
ionization of this molecule has been experimentally shown \cite{Domagoj} to
deviate from the predictions of MO-ADK, a single-active-electron quasi-static
model of molecular ionization.  Our method allows the inclusion of two dominant
effects not present in MO-ADK: 1) influence of the specific shape of the
tunneling barrier discussed in Sec.III-C and 2) the possibility to rigorously
couple multiple ionic channels as dissused (but presently not implementd) in
Sec.III-D.  In our analysis, the deviations from MO-ADK seen experimentally
likely arise from intermediate ionic core excitations followed by interchannel
coupling.

\appendix


\section{Matrix Elements of the Hamiltonian}

\begin{widetext}

In order to evaluate the matrix elements appearing in Eq.(\ref{EqFinalMultiIon}),
we need to know how $ H^F(\{{\vec r}_n\},t)$ acts on the basis states.
The Hamiltonian acting on the neutral state gives
\begin{eqnarray}\label{EqHamilNeutral}
	 H^F |N\rangle &=& \left( E^N_0 - \sum_{k=1}^n \vec F(t)\cdot \vec r_k\right) 
	|N\rangle.
\end{eqnarray}
The Hamiltonian acting on a state $|\phi_m\rangle|I_m\rangle$, where $|I_m\rangle$ is an ionic 
state and $|\phi_m\rangle$ is here an arbitrary single particle function, gives 
\begin{eqnarray}\label{EqHamilIonic}
	 H^F \left( |\phi_m\rangle|I_m\rangle \right)
	&=& \left[  H^I
	-\frac{1}{2}{\vec \nabla}^2_n + V_{nuc}({\vec r}_n)
             +\sum_{k=1}^{n-1}\frac{1}{|{\vec r}_k-{\vec r}_n|} - \sum_{k=1}^n\vec F(t) \cdot \vec r_k
	\right]
	 |\phi_m\rangle|I_m\rangle 
\\ \nonumber
	&=& 
	\sum_j |I_j\rangle \langle I_j|
	 \left[ E^I_m
	-\frac{1}{2}{\vec \nabla}^2_n + V_{nuc}({\vec r}_n)
             +\sum_{k=1}^{n-1}\frac{1}{|{\vec r}_k-{\vec r}_n|} - \sum_{k=1}^n\vec F(t) \cdot \vec r_k
	\right]
	 |\phi_m\rangle|I_m\rangle 
\\ \nonumber
	&=& 
	  \left(  H_m(\vec r_n) |\phi_m\rangle\right)|I_m\rangle
	+ \left( \sum_{j\neq m} V^H_{jm}(\vec r_n)
	|\phi_m\rangle \right) |I_j\rangle 
	- \sum_{j} |I_j\rangle
	\langle I_j| \sum_{k=1}^n \vec F(t)\cdot\vec r_k \left( |\phi_m\rangle|I_m\rangle\right)
\\ \nonumber
	&=& 
	  \left(  H_m(\vec r_n) |\phi_m\rangle\right)|I_m\rangle
	+ \left( \sum_{j\neq m} V^H_{jm}(\vec r_n)
	|\phi_m\rangle \right) |I_j\rangle
	- \vec F(t) \cdot \left(\sum_{j} ( {\vec r}_n\delta_{jm} - {\vec d}^I_{jm})
	|\phi_m\rangle  \right) |I_j\rangle
\end{eqnarray}
where
\begin{equation}
 H_m(\vec r_n) =  E^I_m-\frac{1}{2}{\vec \nabla}^2_n + V_{nuc}({\vec r}_n)
	+ V^H_{mm}({\vec r}_n)  ,
\end{equation}
is the single-electron field-free Hamiltonian for the $n^{th}$ electron 
coupled to the ionic state $|I_m\rangle$,
\begin{equation}
	{\vec d}^I_{jm} = -\langle I_j| \sum_{k=1}^{n-1} \vec r_k  |I_m\rangle
\end{equation}
are the electronic dipole moments and transition dipoles of the ionic states,
and 
\begin{equation}
	V^H_{jm}(\vec r_n) = \langle I_j| \sum_{k=1}^{n-1}\frac{1}{|{\vec r}_k-{\vec r}_n|} |I_m\rangle
\end{equation}
are the electrostatic potentials and inter-ionic couplings.
Eq.~(\ref{EqHamilIonic}) is only exact if a complete basis of $|I_j\rangle$ is used.
If this basis is truncated, Eq.~(\ref{EqHamilIonic}) gives 
$H^F \left( |\phi_m\rangle|I_m\rangle \right)$ projected into the space of the
truncated basis.
Below we will also need the electronic dipole of the neutral defined as
\begin{equation}
	{\vec d}^N = -\langle N| \sum_{k=1}^n \vec r_k  |N\rangle
\end{equation}

We now calculate the required matrix elements of the Hamiltonian. First
First consider the matrix elements of 
the `primitive' basis functions $|N\rangle$, $|S_m\rangle$, and $|X_m(t)\rangle$.
In the following matrix elements the convention $m\neq k$ is used in order to 
avoid excessive use of Kronecker's delta.
\begin{eqnarray}
	\langle S_m|  H^F(t) | S_m\rangle &=&
	(\langle \tilde \phi^S_m|\langle I_m|)  H^F(t) (|\tilde \phi^S_m\rangle|I_m\rangle)
	\nonumber \\
	&=& \langle \tilde \phi^S_m| H_m |\tilde \phi^S_m\rangle
	- \vec F(t) \cdot [ \langle \tilde \phi^S_m|\vec r_n |\tilde \phi^S_m\rangle- \vec d^I_{mm} ]
\end{eqnarray}
\begin{eqnarray}
	\langle S_m|  H^F(t) | S_k\rangle &=&
	(\langle \tilde \phi^S_m|\langle I_m|)  H^F(t) (|\tilde \phi^S_k\rangle|I_k\rangle)
	\nonumber \\
	&=& \langle \tilde \phi^S_m|V^H_{mk} |\tilde \phi^S_k\rangle
	+ \vec F(t) \cdot \vec d^I_{mk} \langle \tilde \phi^S_m|\tilde \phi^S_k\rangle
\end{eqnarray}
\begin{eqnarray}
	\langle S_m|  H^F(t) | N\rangle &=&
	(\langle \tilde \phi^S_m|\langle I_m|)  H^F(t) |N\rangle
    \nonumber \\
	&=& \eta_m E^N_0 
	- \vec F(t) \cdot  [\eta_m\langle \tilde \phi^S_m| \vec r_n |\tilde \phi^S_m\rangle +
	\langle \tilde \phi^S_m|\vec \phi^C_m\rangle]
\end{eqnarray}
\begin{eqnarray}
	\langle S_m|  H^F(t) |X_m(t)\rangle &=&
	(\langle \tilde \phi^S_m|\langle I_m|)  H^F(t) (|\chi_m(t)\rangle|I_m\rangle)
	\nonumber \\
	&=& \langle \tilde \phi^S_m| H_m |\chi_m(t)\rangle
	- \vec F(t) \cdot  \langle \tilde \phi^S_m|\vec r_n |\chi_m(t)\rangle
\end{eqnarray}
\begin{eqnarray}
	\langle S_m|  H^F(t) |X_k(t)\rangle &=&
	(\langle \tilde \phi^S_m|\langle I_m|)  H^F(t) (|\chi_k(t)\rangle|I_k\rangle)
	\nonumber \\
	&=& \langle \tilde \phi^S_m|V^H_{mk} |\chi_k(t)\rangle
	+ \vec F(t) \cdot \vec d^I_{mk} \langle \tilde \phi^S_m|\chi_k(t)\rangle 
\end{eqnarray}
\begin{eqnarray}
	\langle N|  H^F(t) | N\rangle &=&
	E^N_0 + \vec F(t) \cdot \vec d^N
\end{eqnarray}
\begin{eqnarray}
	\langle N|  H^F(t) |X_m(t)\rangle &=&
	\langle N|  H^F(t) (|\chi_m(t)\rangle|I_m\rangle)
	\nonumber \\
	&=& - \vec F(t) \cdot [\eta_m^*\langle \tilde \phi^S_m|\vec r_n |\chi_m(t)\rangle
	+ \langle \vec \phi^C_m|\chi_m(t)\rangle]
\end{eqnarray}
\begin{eqnarray}
	  \langle I_m|  H^F(t) | S_m\rangle  &=&
	  \langle I_m|  H^F(t) (|\tilde \phi^S_m\rangle|I_m\rangle)
	 \nonumber \\
	&=&
	   H_m |\tilde \phi^S_m\rangle
	- \vec F(t) \cdot ( \vec r_n - \vec d^I_{mm}) |\tilde \phi^S_m\rangle 
\end{eqnarray}
\begin{eqnarray}
	  \langle I_m|  H^F(t) | S_k\rangle  &=&
	  \langle I_m|  H^F(t) (|\tilde \phi^S_k\rangle|I_k\rangle)
	 \nonumber \\
	&=&
	  V^H_{mk} |\tilde \phi^S_k\rangle
	+ \vec F(t) \cdot \vec d^I_{mk} |\tilde \phi^S_k\rangle 
\end{eqnarray}
\begin{eqnarray}
	  \langle I_m|  H^F(t) | N\rangle  &=&
	  \eta_m E^N_0 |\tilde \phi^S_m\rangle 
	- \vec F(t) \cdot [ \vec r_n |\tilde \phi^S_m\rangle\eta_m + |\vec \phi^C_m\rangle]
\end{eqnarray}
\begin{eqnarray}
	  \langle I_m|  H^F(t) | X_m(t)\rangle  
	&=& \langle I_m|  H^F(t) (|\chi_m(t)\rangle|I_m\rangle) 
	 \nonumber \\
	&=&  
	 H_m |\chi_m(t)\rangle 
	- \vec F(t) \cdot ( \vec r_n - \vec d^I_{mm})|\chi_m(t)\rangle 
\end{eqnarray}
\begin{eqnarray}
	  \langle I_m|  H^F(t) | X_k(t)\rangle  
	&=& \langle I_m|  H^F(t) (|\chi_k(t)\rangle|I_k\rangle) 
	 \nonumber \\
	&=&  
	V^H_{mk} |\chi_k(t)\rangle 
	+ \vec F(t) \cdot \vec d^I_{mk}|\chi_k(t)\rangle 
\end{eqnarray}

Now we use these matrix elements to evaluate the remaining terms in Eqs.~(\ref{EqFinalMultiIon}) 
that involve $|\tilde N\rangle$ 
\begin{eqnarray}
	\langle S_m|  H^F(t) |\tilde N\rangle &=&
	{\cal N}_{\tilde N} \Big[ \langle S_m|  H^F(t) |N\rangle
	-\sum_k\eta_k\langle S_m|  H^F(t) | S_k \rangle \Big]
\\ \nonumber
	&=&
	{\cal N}_{\tilde N}\Big[ \langle S_m|  H^F(t) |N\rangle
	-\eta_m\langle S_m|  H^F(t) | S_m \rangle
	-\sum_k^{k\neq m}\eta_k\langle S_m|  H^F(t) | S_k \rangle \Big]
\\ \nonumber
	&=& {\cal N}_{\tilde N} \Big[
	\eta_m E^N_0
    - \vec F(t) \cdot  [\eta_m\langle \tilde\phi^S_m|\vec r_m|\tilde\phi^S_m\rangle  +
    \langle \tilde \phi^S_m|\vec \phi^C_m\rangle]
	\\ \nonumber
	&-&\eta_m[ 
	 \langle\tilde\phi^S_m| H_m|\tilde\phi^S_m\rangle
	- \vec F(t) \cdot [ \langle \tilde\phi^S_m|\vec r_m|\tilde\phi^S_m\rangle - \vec d^I_{mm} ]
	]
	\\ \nonumber
	&-&
    \sum_k^{k\neq m}\eta_k [ \langle\tilde\phi^S_m|V^H_{mk}|\tilde\phi^S_k\rangle
    + \vec F(t) \cdot \vec d^I_{mk} \langle \tilde \phi^S_m|\tilde \phi^S_k\rangle ]
	\Big]
\\ \nonumber
	&=& {\cal N}_{\tilde N} \Big[
	\eta_m [E^N_0 - \langle\tilde\phi^S_m| H_m|\tilde\phi^S_m\rangle ]
    - \vec F(t) \cdot  [
    \langle \tilde \phi^S_m|\vec \phi^C_m\rangle
	+ \eta_m \vec d^I_{mm} ]
	\\ \nonumber
	&-&
    \sum_k^{k\neq m}\eta_k [ \langle\tilde\phi^S_m|V^H_{mk}|\tilde\phi^S_k\rangle
    + \vec F(t) \cdot \vec d^I_{mk} \langle \tilde \phi^S_m|\tilde \phi^S_k\rangle ]
	\Big]
\end{eqnarray}
\begin{eqnarray}
	\langle \tilde N|  H^F(t) | S_m\rangle &=&
	\langle S_m|  H^F(t) |\tilde N\rangle^* 
	\\ \nonumber
	&=& {\cal N}_{\tilde N} \Big[
	\eta_m^* [E^N_0 - \langle\tilde\phi^S_m| H_m|\tilde\phi^S_m\rangle ]
    - \vec F(t) \cdot  [
    \langle \vec \phi^C_m|\tilde \phi^S_m\rangle
	+  \eta_m^* \vec d^I_{mm} ]
	\\ \nonumber
	&-&
    \sum_k^{k\neq m}\eta_k^* [ \langle\tilde\phi^S_k|V^H_{km}|\tilde\phi^S_m\rangle
    + \vec F(t) \cdot \vec d^I_{km} \langle \tilde \phi^S_k|\tilde \phi^S_m\rangle ]
	\Big]
\end{eqnarray}
\begin{eqnarray}
	\langle \tilde N|  H^F(t) |\tilde N\rangle 
	&=& |{\cal N}_{\tilde N}|^2  \left(\langle N|-\sum_m\eta^*_m\langle S_m|\right) 
	 H^F(t) \left(|N\rangle - \sum_k\eta_k| S_k\rangle\right)
\nonumber \\
	&=&  |{\cal N}_{\tilde N}|^2 \langle N|  H^F(t) |N\rangle 
	    +|{\cal N}_{\tilde N}|^2\sum_{mk}^{k\neq m}\eta_m^*\eta_k\langle S_m|  H^F(t) | S_k\rangle
     \nonumber \\
	    &+& |{\cal N}_{\tilde N}|^2\sum_m\left[ |\eta_m|^2 \langle S_m|  H^F(t) | S_m\rangle 
		-\eta_m\langle N|  H^F(t)  | S_m\rangle 
	    -\eta^*_m\langle S_m| H^F(t) |N\rangle \right]
\end{eqnarray}
\begin{eqnarray}
	\langle \tilde N|  H^F(t) |X_m(t)\rangle &=&
	{\cal N}_{\tilde N} \Big[ \langle N|  H^F(t) |X_m(t)\rangle
	-\sum_k\eta^*_k\langle S_k|  H^F(t) |X_m(t)\rangle
	\Big]
    \nonumber \\
    &=& {\cal N}_{\tilde N} \Big[
    - \vec F(t) \cdot [\eta_m^* \langle \tilde \phi^S_m|\vec r_n |\chi_m(t)\rangle
    + \langle \vec \phi^C_m|\chi_m(t)\rangle]
	\nonumber \\
    &-& \eta^*_m [ \langle\tilde \phi^S_m| H_m |\chi_m(t)\rangle
    - \vec F(t) \cdot  \langle \tilde\phi^S_m|\vec r_n | \chi_m(t) \rangle
    ]
	\nonumber \\ 
    &-& \sum_k^{k\neq m} \eta^*_k [
	 \langle \tilde \phi^S_m|V^H_{mk} |\chi_k(t)\rangle
	+ \vec F(t) \cdot \vec d^I_{mk} \langle \tilde \phi^S_m|\chi_k(t)\rangle 
    ]
	\Big]
    \nonumber \\
    &=&{\cal N}_{\tilde N} \Big[
    -\eta^*_m \langle \tilde \phi^S_m| H_m |\chi_m(t)\rangle
    - \vec F(t) \cdot  \langle \vec \phi^C_m|\chi_m(t)\rangle
	\nonumber \\ 
    &-& \sum_k^{k\neq m} \eta^*_k [
	 \langle \tilde \phi^S_m|V^H_{mk} |\chi_k(t)\rangle
	+ \vec F(t) \cdot  \vec d^I_{mk} \langle \tilde \phi^S_m|\chi_k(t)\rangle 
    ]
	\Big]
\end{eqnarray}
\begin{eqnarray}
	\langle I_m|  H^F(t) | \tilde N\rangle 
	 &=&{\cal N}_{\tilde N} \Big[
	  \langle I_m|  H^F(t) | N\rangle 
	  - \sum_k \eta_k \langle I_m|  H^F(t) | S_k\rangle 
	\Big]
\\ \nonumber
	 &=& {\cal N}_{\tilde N} \Big[
	  E^N_0 |\tilde \phi^S_m\rangle \eta_m
	- \vec F(t) \cdot  \vec r_n |\tilde \phi^S_m\rangle\eta_m  
	- \vec F(t) \cdot|\vec \phi^C_m\rangle
	\\ \nonumber
	 &-& \eta_m 
	   H_m |\tilde \phi^S_m\rangle
	+ \eta_m \vec F(t) \cdot ( \vec r_n - \vec d^I_{mm}) |\tilde \phi^S_m\rangle 
	\\ \nonumber
	&-&\sum_k^{k\neq m} \eta_k \Big[
	  V^H_{mk} |\tilde \phi^S_k\rangle
	+ \vec F(t) \cdot \vec d^I_{mk} |\tilde \phi^S_k\rangle 
	\Big]
	\Big]
\\ \nonumber
	 &=&{\cal N}_{\tilde N} \Big[
	  \eta_m [ E^N_0 - \vec F(t) \cdot  \vec d^I_{mm} ]
	|\tilde \phi^S_m\rangle 
	 - \eta_m  H_m |\tilde \phi^S_m\rangle
	- \vec F(t) \cdot|\vec \phi^C_m\rangle
	\\ \nonumber
	&-&\sum_k^{k\neq m} \eta_k \Big[
	  V^H_{mk} |\tilde \phi^S_k\rangle
	+ \vec F(t) \cdot \vec d^I_{mk} |\tilde \phi^S_k\rangle \Big]
	\Big]
\end{eqnarray}

\end{widetext}


\end{document}